\documentclass[aps,twocolumn,showpacs,groupedaddress]{revtex4}
\usepackage{graphicx}

\begin{document}

%\draft
\title{Dissipation Effects in Hybrid Systems}
\author{X. X. Yi and W. Wang}
\affiliation{ Department of Physics, Dalian University of
Technology, Dalian 116024, China}

\date{\today}

\begin{abstract}
The dissipation effect in a hybrid system is studied in this
Letter. The hybrid system is a compound of a classical magnetic
particle and a quantum single spin. Two cases are considered. In
the first case, we investigate the effect of the dissipative
quantum subsystem on the motion of its classical partner. Whereas
in the second case we show how the dynamics of the quantum single
spin are affected by the dissipation of the classical particle.
Extension to general dissipative hybrid systems is discussed.
\end{abstract}

 \maketitle

Quantum and classical theories are distinguished both in terms of
their state spaces and their dynamics. Quantum states can predict
measurement results that can not be reconciled with predictions by
classical states, such as violations of Bell's
inequalities\cite{bell64}. Dynamically,  although quantum and
classical evolution agree on sufficiently short time
scales\cite{sakurai93}, the mean values of observables  diverge
after some characteristic time\cite{berry79}. When they go to
dissipative systems, quantum open systems may be described by the
master equation, while the dissipative force proportional to the
momentum of the particle may be introduced to obtain the equations
of motion for dissipative classical systems. Then a question
arises, in a hybrid system composed of a quantum subsystem and a
classical subsystem, how to treat the dissipation effects? And
what are the  effects? This question became more important in the
last years because of remarkable progress\cite{koppens05,
yamamoto03} made in experiments in quantum information processing,
where the qubits have to be coupled to the macroscopic world for
initialization, gating and readout. For example, in a typical flux
qubit gate, microwave pulses are applied onto specific qubit of
the sample. This requires many classical systems coupling to the
qubits, which is thus a compound of quantum and classical
subsystems.

Besides its experimental interest, it is of central importance on
the scientific side. For a closed hybrid system, the quantum
subsystem may be treated classically \cite{slichter90,berman02} or
quantum mechanically\cite{zhang06}, depending on specific issues
addressed. The formulism requires that the hybrid system can be
described by a Hamiltonian. This requirement, however, is not
feasible for subsystems that include dissipation, in particular,
for the quantum dissipative subsystem. The so-called
system-plus-reservoir approach is to consider the classical
subsystem as a reservoir, leading to decoherence in the quantum
subsystem. This approach, however, ignore the backaction of the
quantum subsystem, and hence is inadequate to address some issues,
for instance, in single spin detection by the
magnetic-resonance-force microscopy\cite{rugar04}.

\begin{figure}
\includegraphics*[width=0.95\columnwidth,
height=0.4\columnwidth]{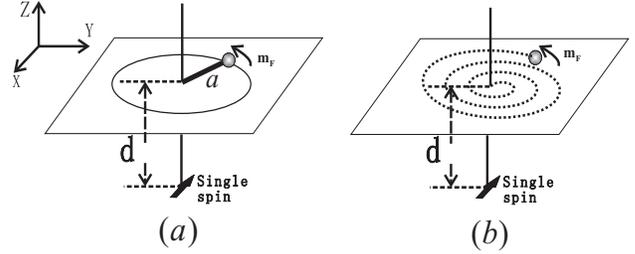} \caption{A schematic
illustration of a single spin-$\frac 1 2 $ particle interacting with
a magnetic particle. ($a$) The magnetic particle is attached to a
rigid cantilever and rotates  in the $xy$ plane. The single spin
placed beneath the plane with a distance of $d$ is dissipative due
to coupling to its environment. ($b$) The magnetic particle moves
dissipatively in the $xy$ plane, while the spin placed beneath the
plane is decoherence free.} \label{fig1}
\end{figure}
In this Letter, we present a method for dissipative hybrid
systems. The hybrid system consists of a quantum subsystem, which
is dynamically fast, and a slow classical subsystem. The presented
representation is based on the fact that a quantum system
possesses mathematically a canonical classical Hamiltonian
structure\cite{heslot85, weinberg89, liu03}. In fact, this method
was used in closed hybrid systems in \cite{zhang06}. To show the
dissipation effects, we first consider the case where the quantum
subsystem includes dissipation, while the classical subsystem does
not dissipate. We examine the effect of the dissipative quantum
system on the motion of the classical subsystem. The second case
we consider is that only the classical subsystem in the hybrid
system is dissipative. The dynamics as well as the adiabaticity of
the quantum subsystem are examined. We find that large dissipation
rates of the classical subsystem benefit the adiabatic evolution
of the quantum subsystem, and the vector potential arised in the
hybrid system tends to zero with the dissipation rate approaching
infinity. On the other hand, the dissipative quantum subsystem
produces a magnetic-like field  in the slow classical subsystem.
The strength of this field oscillates at intermediate values of
the dissipation rate, and then approaches zero with large
dissipation rates.

We shall use a simple hybrid system, which consists of a single
spin coupling to a heavy magnetic particle, to show the idea.
The assumption that
 the magnetic particle is dynamically slow with respect to the single spin
  is of relevance to the single spin
detection by the magnetic-resonance-force microscopy\cite{rugar04},
where  the cantilever plays the role of the slow classical
subsystem. A schematic setup of the studied system was shown in
figure 1. A magnetic particle with magnetic moment, ${\bf m}_F$, is
attached to the cantilever tip (Fig. 1-(a)). The cantilever is rigid
and rotates freely in the $xy$ plane. A single spin with magnetic
moment $\vec{\mu}$ is placed beneath the plance with a distance of
$d$. We shall show the effect of the dissipative quantum subsystem
on the motion of the magnetic particle through this imagined setup.
Whereas by the setup   in Fig. 1-(b), we shall study the dissipation
effect of the classical subsystem on the quantum subsystem. The
difference of the two setups is that, in the setup in Fig. 1-(b),
the magnetic particle may move freely in the $xy$ plane, i.e., there
is not any rigid cantilever for the magnetic particle to be
attached.

Consider the hybrid system sketched in Fig. 1-(a). The dynamics of
the single spin-$\frac 1 2$ particle can be described by the
master equation,
\begin{equation}
\dot{\rho}=-i[\hat{H}(\vec{q}_b),\rho]+\frac{\gamma}{2}\{
2\sigma_-\rho\sigma_+-\rho\sigma_+\sigma_--\sigma_+\sigma_-\rho\},
\label{mse}
\end{equation}
where $\hat{H}(\vec{q}_b)=\mu \vec{B}(\vec{q}_b)\cdot
\vec{\sigma}$ denotes the system Hamiltonian of the quantum single
spin, $\gamma$ stands for the spontaneous emission rate.
$\sigma_i, i=z,+,- $ are the Pauli matrices, and
$\sigma_z=|e\rangle\langle e|-|g\rangle\langle g|$ ($|e\rangle$,
denotes the  state of spin-up, and $|g\rangle$ spin-down). We
shall denote $H_b(\vec{p}_b, \vec{q}_b)$ the Hamiltonian of the
heavy classical subsystem $b$ that is dynamically slow, and
$\vec{p}_b, \vec{q}_b$ are its momenta and coordinates,
respectively.  The dependence of $\hat{H}_a(\vec{q}_b)$ on
$\vec{q}_b$ indicates the coupling between the two subsystems. The
magnetic field  acting on the single spin from the classical
magnetic particle is given by ($\vec{q}_b=\{ x, y, d\} $)
\cite{zhang06}
\begin{equation}
\vec{B}(\vec{q}_b)=\{ B_x, B_y, B_z\}=-\frac{\mu_0m_F\{ 3dx, 3dy,
2d^2-a^2\}}{4\pi(d^2+a^2)^{5/2}},
\end{equation}
where $a=\sqrt{x^2+y^2}$ keeps fixed in this case.   The key idea
required to use the framework\cite{zhang06,heslot85, weinberg89,
liu03} for the exact treatment of the hybrid system is to find an
effective Hamiltonian for the quantum subsystem. The method for
this purpose was first presented in \cite{yi01}, the idea is the
following. The density matrix $\rho(t)$ of the open system can be
mapped onto a pure state by introducing an ancilla. The dynamics
of the open system is then described by a Schr\"odinger-like
equation with an effective Hamiltonian that can be derived from
the master equation. In this way the solution of the master
equation can be obtained in terms of the evolution of the
composite system by converting the pure state back to the density
matrix.  For the single spin particle under consideration, we may
introduce the other spin-$\frac 1 2 $ particle with energy levels
labelled by $|e\rangle^A$ and $|g\rangle^A$ as the ancilla. In
spirit of the effective Hamiltonian approach, a pure state for the
composite system (the single spin plus the ancilla) may be
constructed as
\begin{equation}
|\Psi_{\rho}(t)\rangle=\sum_{m,n=e,g}\rho_{mn}(t)|m\rangle|n\rangle^A,\label{ee1}
\end{equation}
where $\rho_{mn}(t)$ are density matrix elements of the open
system in the  basis $\{|e\rangle,|g\rangle \}$, i.e.,
$\rho_{mn}(t)=\langle m|\rho(t)|n\rangle, m,n=e,g. $  With these
notations, we may find an effective Hamiltonian ${\cal
H}_T(\vec{q}_b)$, such that the bipartite(the spin plus the
ancilla) pure state $|\Psi_{\rho}(t)\rangle$ satisfies the
following Schr\"odinger-like equation
\begin{equation}
i\frac{\partial}{\partial t}|\Psi_{\rho}(t)\rangle={\cal
H}_T(\vec{q}_b)|\Psi_{\rho}(t)\rangle.\label{se1}
\end{equation}
  To shorten  the derivation, we write the
master equation Eq.(\ref{mse}) as
\begin{equation}
i\frac{\partial}{\partial t}\rho(t) = {\cal H}\rho(t)
-\rho(t){\cal H}^{\dagger}
 +i\gamma\sigma_-\rho(t)\sigma_+, \label{me2}
\end{equation}
with ${\cal H}
=\hat{H}_a(\vec{q}_b)-i\frac{\gamma}{2}|e\rangle\langle e|.$
Substituting equation Eq.(\ref{ee1}) together with Eq.(\ref{me2})
into Eq.(\ref{se1}), one finds after some algebra,
\begin{equation}
{\cal H}_T(\vec{q}_b)={\cal H} -{\cal H}^A
+i\gamma\sigma_-^A\sigma_-.
\end{equation}
Operators ${\cal H} $ and $\sigma_-$ are for the single spin,
which take the same form as in Eq.(\ref{me2}), while ${\cal H}^A$
and $\sigma_-^A$ are operators for the ancilla  defined by
\begin{equation}
^A\langle m|\hat{O}^A|n \rangle^A=\langle n| \hat{O}^{\dagger} |m
\rangle, m, n=e,g,
\end{equation}
with $ \hat{O} ={\cal H},$  or $\sigma_-.$ This yields ${\cal H}^A
(\vec{q}_b)=\hat{H}^A(\vec{q}_b)+\frac i 2|e\rangle^{AA}\langle
e|,$ $\hat{H}^A(\vec{q}_b) =\mu \vec{B}(\vec{q}_b) \cdot
\vec{\sigma}^A$, and $\vec{\sigma}^A $ represents the Pauli matrix
of the ancilla. The first two terms in the effective Hamiltonian
${\cal H}_T(\vec{q}_b)$ describe the free evolution of the spin
and   ancilla, respectively, and the third term characterizes
couplings between the spin and the ancilla.

When the quantum subsystem is dynamically fast and the classical
subsystem is slow, a vector potential $\vec{A}$  is generated in
the hybrid system \cite{zhang06}. This vector potential behaves
like the familiar Berry phase in the quantum subsystem, while it
enters the classical subsystem in terms of magnetic-like fields
$\vec{b}=\nabla\times \vec{A}$. With these knowledge, the total
Hamiltonian for the dissipative hybrid system can be expressed in
a pure classical formulism,
\begin{equation}
\tilde{H}={\cal H}_T(\vec{I}_a, \vec{q}_b)+H_b(\vec{P}_a-\vec{A},
\vec{q}_b),
\end{equation}
where ${\cal H}_T(\vec{I}_a, \vec{q}_b)=\sum_{i=1}^4\lambda_ip_i$,
 $(\vec{I}_b)_i=\hbar p_i,$
  $H_b(\vec{P}_a-\vec{A}, \vec{q}_b)$ stands for the  Hamiltonian of the
classical subsystem, $\lambda_i$ $(i=1,2,3,4)$ are the eigenvalues
of the effective Hamiltonian ${\cal H}_T$, and  $p_j$ is the
probability of finding the bipartite  system (the single spin plus
the ancilla) on state \cite{yi06},
\begin{equation}
|\psi_j\rangle=\frac{1}{\sqrt{M_j}}(a_j|e\rangle|e\rangle^A
+b_j|e\rangle|g\rangle^A+c_j|g\rangle|e\rangle^A+d_j|g\rangle|g\rangle^A).
\end{equation}
Tedious but standard calculations show that under the adiabatic
evolution in open systems\cite{sarandy05}(see also \cite{yi06}) the
vector potential and magnetic-like field are $\vec{A}=\{ A_x,
A_y,0\}$ and $\vec{b}=\{ 0,0,b_z\}$, with ($\lambda_4=0$)
\begin{eqnarray}
A_x&=&-\hbar\sum_{j=1}^3
\frac{p_j}{M_j}(C_jc_j-B_jb_j)\frac{y}{a^2},\nonumber\\
A_y&=&\hbar\sum_{j=1}^3
\frac{p_j}{M_j}(C_jc_j-B_jb_j)\frac{x}{a^2},\nonumber\\
 b_z&=&2\hbar\sum_{j=1,2,3}
\frac{p_j}{a^2 M_j}(C_jc_j-B_jb_j),
\end{eqnarray}
where
\begin{eqnarray}
M_j&=&A_ja_j+B_jb_j+C_jc_j+D_jd_j,\nonumber\\
a_j&=&-(2\cos\theta+\frac 1 2 i\gamma+\lambda_j),\nonumber \\
b_j&=&2\sin\theta e^{i\phi},\nonumber\\
c_j&=&\frac{2\sin\theta(2\cos\theta+\frac 1 2 i\gamma+\lambda_j)
e^{-i\phi}}{2\cos\theta-\frac 1 2 i\gamma-\lambda_j},\nonumber\\
d_j&=&-a_j,\nonumber\\
A_j&=&-(2\cos\theta-\frac 1 2 i\gamma+\lambda_j),\nonumber \\
D_j&=&\frac{i\gamma-\lambda_j}{i\gamma+\lambda_j}A_j,\nonumber\\
B_j&=&\frac{\sin\theta
e^{-i\phi}(D_j-A_j)}{2\cos\theta+\frac 1 2 i\gamma+\lambda_j},\nonumber\\
C_j&=&\frac{\sin\theta e^{i\phi}(D_j-A_j)}{2\cos\theta-\frac 1 2
i\gamma-\lambda_j},\nonumber\\
\cos\theta&=&\frac{B_z}{|\vec{B}|}, \ \ \tan\phi=\frac{B_y}{B_x},
\end{eqnarray}
and $\lambda_j$ is given by,
 \begin{eqnarray}
(\lambda_j&+&\frac i 2 \gamma)^3+
\frac i 2\gamma(\lambda_j+\frac i 2\gamma)^2-4(\lambda_j
+\frac i 2\gamma)\nonumber\\
&=&2i\gamma^2\cos^2\theta.
\end{eqnarray}
We choose $p_1=p_2=p_3=\frac 1 3$ to show the dependence of
$Re(b_z)$ (the real part of $b_z$)  on the dissipation rate
$\gamma$ and the distance $d$. The numerical results were
presented in figure 2. We find from figure 2 that $Re(b_z)$ decays
with $d$ increasing. This is easy to understand, since $d$
represents the distance between the magnetic particle and the
single spin. One can also see from figure 2 that the dependence of
$Re(b_z)$ on $\gamma$ is a oscillating function. It arrives at its
maximum with a nonzero $\gamma$, and then tends to zero with
$\gamma\gg\mu|\vec{B}|.$
\begin{figure}
\includegraphics*[width=0.8\columnwidth,
height=0.6\columnwidth]{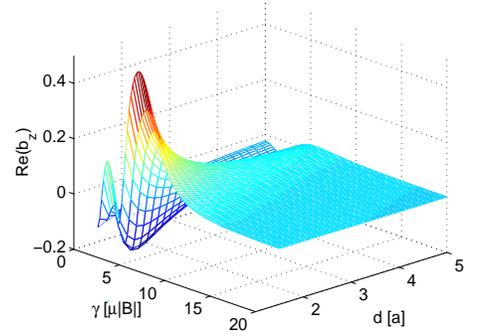} \caption{ Real part of the
magnetic-like field $Re(b_z)$ versus the distance $d$ and the
dissipation rate $\gamma.$ $p_1=p_2=p_3$ were chosen for this
plot, and $Re(b_z)$ was calculated in units of $\sqrt{m\mu
|\vec{B}|}/a$. } \label{fig2}
\end{figure}
Note that the dependence of the vector potential $\vec{A}$ on
$\gamma$ and $d$ is similar to that of the magnetic-like field
$\vec{b}$, hence $\vec{A}$ has the same feature as     we
presented above for $\vec{b}$. With this observation, the equation
of motion for the magnetic particle (mass $m$) takes,
\begin{equation}
m\ddot{\bar{q}}_b=-\sum_j\frac{\partial \lambda_j}{\partial\vec{
q}_b}p_j-m\Omega^2 \vec{q}_b+\dot{\vec{q}}_b\times
\vec{b},\label{eofm}
\end{equation}
where $\Omega$ is the frequency with that the classical particle
circles. The first term on the right-hand side in Eq.(\ref{eofm})
may be set to zero by properly choosing $p_j$. In this situation,
the dissipation effect enters the classical particle through the
magnetic-like field, and causes damping in the classical
particle's motion.

Now, we turn to study the second case illustrated in Fig.1-(b). In
this case, the quantum subsystem is decoherence free and described
by $\hat{H}_a(\vec{q}_b)=\mu\vec{B}(\vec{q}_b)\cdot\vec{\sigma},$
  but the classical particle is
subject to dissipations. We introduce a dissipative force $2\Gamma
\dot{\vec{q}_b}$ to obtain the equations of motion in the form
\begin{eqnarray}
\dot{\vec{p}_b}&=&-m\Omega^2\vec{q}_b-2\Gamma\frac{\vec{p}_b}{m},\nonumber\\
\dot{\vec{q}_b}&=&\frac{\vec{p}_b}{m}.
\end{eqnarray}
This yields the  well known damped solution
\begin{equation}
\vec{q}_b=e^{-\frac{\Gamma}{m}t}(\vec{A}\cos(\Omega_0
t)+\vec{B}\sin(\Omega_0t)),\label{co}
\end{equation}
where $\Omega_0=\sqrt{\Omega^2-(\Gamma/m)^2}$, $\vec{A}$ and
$\vec{B}$ depend on the initial condition of the magnetic
particle. We are interested in the dynamics of the quantum
subsystem under the influence of the damped magnetic particle.
First, we study the effect of dissipation on the adiabaticity of
the single spin. Assuming that $\vec{q}_b$ are just some fixed
parameters, we obtain the instantaneous eigenstates for the
quantum subsystem,
$|+\rangle=\sin(\theta/2)e^{i\phi}|g\rangle+\cos(\theta/2)|e\rangle,$
and
$|-\rangle=\cos(\theta/2)e^{i\phi}|g\rangle-\sin(\theta/2)|e\rangle,$
with $\cos\theta=B_z/|\vec{B}|,$ and $\tan\phi=B_y/B_x.$   The
corresponding eigenenergies  are $E_+=\mu|\vec{B}|,$
$E_-=-\mu|\vec{B}|,$  respectively.   The adiabatic evolution of
the single spin requires that
\begin{equation}
\kappa(\Gamma, t)=\frac 1 4 |\dot{\theta}+i\Omega_0\sin\theta|\ll
1.
\end{equation}
For $\Gamma=0,$ i.e., the magnetic particle is not dissipative,
$\kappa(\Gamma,t)$ reduces to $\kappa(\Gamma, t)=\frac 1 4
\Omega|\sin\theta|$, it depends on the frequency $\Omega$  and
radius with that the magnetic particle circles. The dependence of
$\kappa(\Gamma, t)$ on the damping rate $\Gamma$ and time $t$ was
presented in figure 3.
\begin{figure}
\includegraphics*[width=0.8\columnwidth,
height=0.6\columnwidth]{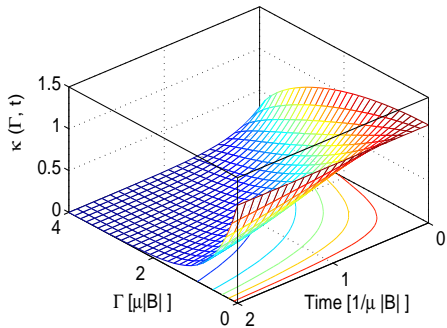} \caption{ $\kappa(\Gamma,
t)$ as a function of dissipation rate and time $t$.
$\kappa(\Gamma, t)$ can be used to characterize the adiabaticity
of the quantum subsystem. The parameters chosen are the same as in
figure 2.} \label{fig3}
\end{figure}
One can see from figure 3 that large dissipation rate $\Gamma$
would benefit the adiabatic evolution of the single spin. This can
be understood by examining the effective frequency $\Omega_0$.
Clearly, the larger the $\Gamma$, the smaller the $\Omega_0.$ From
figure 3 we can also find that with time evolution, the adiabatic
condition becomes easier to meet, this can be interpreted as the
slowdown of  the moving magnetic particle.  Next, we study
numerically the dissipation effect on  the time evolution of the
single spin, by calculating numerically the population on the
state $|e\rangle.$ Extensive numerical simulations with the
Hamiltonian $H_a(\vec{q}_2)$ and the solution Eq. (\ref{co}) have
been performed, we find that for some fixed time points the
population is a decay function of $\Gamma$, but for the other time
points, the population increases first then decays (see figure 4).
Note that the populations in the flat  region in figure 4 are not
zero. This damping effect revealed here is somewhat reminiscent of
the quench effect in Landau-Zener like problems reported recently
in \cite{damski}.
\begin{figure}
\includegraphics*[width=0.8\columnwidth,
height=0.6\columnwidth]{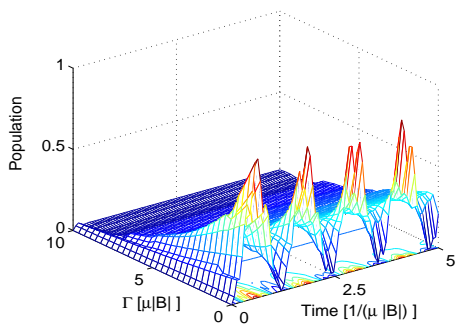} \caption{Population of the
quantum subsystem on state $|e\rangle$. We have chosen $|g\rangle$
as the initial state for this plot. } \label{fig4}
\end{figure}

The  presented   representation for the simple hybrid system can
be easily extended to  general hybrid systems. For an general open
quantum subsystem, the  dynamics may be described  by the master
equation,
\begin{eqnarray}
i\frac{\partial}{\partial
t}\rho(t)&=&[H^{\prime}(\vec{q}_b),\rho(t)] -\frac i 2\sum_k
\{L_k^{\dagger}(\vec{q}_b)L_k(\vec{q}_b)\rho(t)
\nonumber\\
&+&\rho(t)L_k^{\dagger}(\vec{q}_b)L_k(\vec{q}_b)-
2L_k(\vec{q}_b)\rho(t)L_k^{\dagger}(\vec{q}_b)\},\nonumber\\
\label{mefinal}
\end{eqnarray}
where $H^{\prime}(\vec{q}_b)$ is a Hermitian Hamiltonian and
$L_k(\vec{q}_b)$ may be $\vec{q}_b$-dependent operators describing
the system-environment interaction. The same procedure yields the
effective Hamiltonian for the open quantum subsystem,
\begin{equation}
{\cal H}_T^{\prime}(\vec{q}_b)={\cal H}^{\prime}(\vec{q}_b)-{\cal
H}^{\prime,A}(\vec{q}_b)+i\sum_k L_k^A(\vec{q}_b) L_k(\vec{q}_b),
\end{equation}
Where the operators with index $A$ are for the ancilla, which have
the same definition as  given above. In this way, we can discuss
the dissipation effects in this hybrid system as we did in this
Letter.

In conclusion, we have presented a first attempt at studying  the
dissipation effect in hybrid systems. The main results have been
shown through a simple hybrid system, i.e.,  a compound of a
classical magnetic particle and a quantum single spin. On one
hand, the dissipative quantum subsystem   affects the motion of
its classical partner via magnetic-like fields. On the other hand,
the damped classical particle changes the adiabaticity and the
dynamics of the quantum single spin. The method presented here can
be  extended to general hybrid systems readily.

\ \ \\
We thank Biao Wu for helpful discussions. This work was supported
by EYTP of M.O.E, NSF of China (10305002
and 60578014).\\

\end{document}